\documentclass[twocolumn,superscriptaddress,aps]{revtex4}
\usepackage{dcolumn}
\usepackage{bm}
\usepackage{array}
\usepackage{graphicx}
\usepackage{amsmath}
\usepackage{amssymb}
\usepackage{color}

\begin{document}

\title{Large quantum superpositions of a levitated nanodiamond through
spin-optomechanical coupling}
\author{Zhang-qi Yin}
\thanks{yinzhangqi@mail.tsinghua.edu.cn}
\affiliation{The Center for Quantum Information, Institute for Interdisciplinary
Information Sciences, Tsinghua University, Beijing 100084, P. R. China}
\author{Tongcang Li}
\author{Xiang Zhang}
\affiliation{NSF Nanoscale Science and Engineering Center, 3112 Etcheverry Hall,
University of California, Berkeley, California 94720, USA}
\author{L. M. Duan}
\affiliation{The Center for Quantum Information, Institute for Interdisciplinary
Information Sciences, Tsinghua University, Beijing 100084, People's Republic of China}
\affiliation{Department of Physics, University of Michigan, Ann Arbor, Michigan 48109, USA}

\begin{abstract}
We propose a method to generate and detect large quantum superposition states and arbitrary Fock states for the oscillational mode of an optically levitated nanocrystal diamond. The nonlinear interaction required for the generation of non-Gaussian quantum states is enabled through the spin-mechanical coupling with a built-in nitrogen-vacancy center inside the nanodiamond. The proposed method allows the generation of large superpositions of nanoparticles with millions of atoms and the observation of the associated spatial quantum interference under reasonable experimental conditions.
\end{abstract}

\date{\today}
\maketitle
\section{Introduction}
Creating spatial quantum superpositions with massive objects is one of the
most challenging and attractive goals in macroscopic quantum mechanics
\cite{Hornberger12,Bassi13,Nimmrichter13,Chen13}. It provides potential
opportunities to experimentally test different  wave-function collapse models
\cite{Bassi13}, including gravity-induced state reduction \cite{Penrose1996},
which is a manifestation of the apparent conflict between general relativity
and quantum mechanics. Quantum superpositions and interferences have been
realized with electrons, neutrons, atoms, and complex molecules with several
hundred atoms \cite{Schrodinger}. Among different optomechanical systems
\cite{Aspelmeyer13,mechanicalground,Wilson07,Marquardt07}, optically levitated
dielectric particles in vacuum \cite{Chang10,Isart10,li10,li11,Isart11,Isart11b,
Gieseler12,Monteiro13,Pflanzer12,Cheung12,Kiesel13} are particularly promising
for creating  superposition states with the largest \emph{macroscopicity} (as
defined in Ref. \cite{Nimmrichter13}). Due to the absence of the mechanical
contact in this system, the decoherence \cite{Yin09a} can be negligible and the
oscillation frequency is fully tunable. Once cooled to the quantum regime,
optically trapped nanoparticles in vacuum will be ultra-sensitive detectors
\cite{Geraci10,Yin11,Arvanitaki13}, and can be even used to study self-assembly
of the nanoparticles in vacuum \cite{Lechner12,Habraken13}. To generate spatial
quantum superpositions and other non-Gaussian states with an optical cavity,
however, requires a very strong quadratic coupling \cite{Thompson08, Isart11}.
This is a very demanding requirement. To enhance the quadratic coupling, Romero-Isart
\emph{et al.} \cite{Isart11} proposed to prepare spatial quantum superpositions
of nanoparticles with two inter-connected high-finesse optical cavities: one
cavity for cooling, and the other cavity for preparing the superposition state
with a squared position measurement when the nanoparticle falls through it.

In this paper, we propose a scheme to generate and detect arbitrary Fock
states and spatial quantum superposition states for the center-of-mass oscillation
of an optically trapped nanocrystal diamond using the induced
spin-opto-mechanical coupling. The nanodiamond has build-in nitrogen-vacancy
(NV) centers, and electron spins associated with diamond NV centers make
good qubits for quantum information processing as they have nice coherence
properties even at room temperature \cite{Warchtrup06}.
With assistance of a strong magnetic
field gradient from a nearby magnetic tip, strong coupling between the NV\
spin and the mechanical oscillation of the nanodiamond can be achieved.
Using this coupling, we
show how to generate arbitrary Fock states, entangled states, and large
quantum superpositions for the nanodiamond.
 The generated spatial superposition  states and other mesoscopic
quantum superposition states can be detected through different spatial
interference patterns of the nanodiamond.

\begin{figure}[tbph]
\centering
\includegraphics[width=8cm]{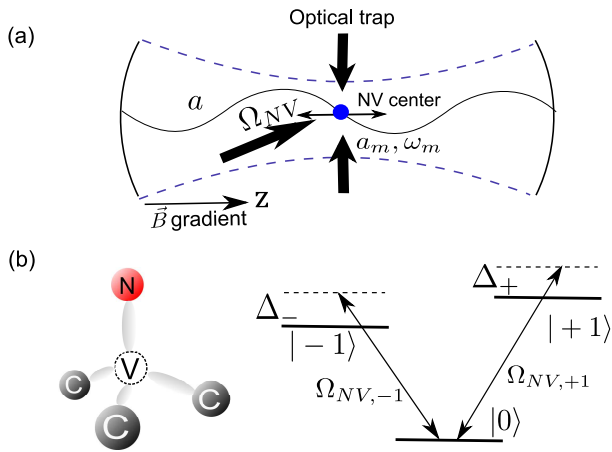}
\caption{(Color online) (a) A nanodiamond with a NV center is optically
trapped in vacuum with spin-mechanical coupling enabled through a nearby
magnetic tip and opto-mechanical coupling through a cavity around. (b) The
atomic structure (left) and the level diagram (right) in the ground state
manifold for a NV center in the nanodiamond . }
\label{fig:scheme}
\end{figure}

\section{The proposed setup}
Nanodiamonds with NV centers have been
recently trapped by optical tweezers in liquid \cite%
{Horowitz12,Geiselmann13} and atmospheric air \cite{Neukirch13}, and similar technologies can be used to optically
trap them in vacuum \cite{Gieseler12}. As shown
in Fig. 1, we consider a nanodiamond of mass $m$ optically trapped in vacuum
with trapping frequency $\omega _{m}$. The motion of its center-of-mass mode
$a_{m}$ is described by the Hamiltonian $H_{m}=\hbar \omega
_{m}a_{m}^{\dagger }a_{m}$. The nanodiamond has a built-in NV center with
its level configuration shown in Fig.1b in the ground-state manifold. The NV
spin is described by the Hamiltonian $H_{\mathrm{NV}}=\hbar (\omega
_{+1}|+1\rangle \langle +1|+\omega _{-1}|-1\rangle \langle -1|)$ , where we
have set $|0\rangle $ as the energy zero point. A magnet tip near the NV
center induces a strong magnetic field gradient \cite{Tsang06,Mamin07}, which
couples the electron spin and the center-of-mass oscillation of the
nanodiamond. The coupling Hamiltonian is denoted by $H_{\mathrm{NV}m}=\hbar
\lambda S_{z}(a_{m}+a_{m}^{\dagger })$ \cite{Rabl09,Kolk12}, where $%
S_{z}\equiv |+1\rangle \langle +1|-|-1\rangle \langle -1|$. The coupling
strength $\lambda =g_{s}\mu _{B}G_{m}a_{0}/\hbar $, where $a_{0}=\sqrt{\hbar
/2m\omega _{m}}$, $g_{s}\simeq 2$ is the Land\'{e} g-factor, $\mu _{B}$ is
the Bohr magneton, and $G_{m}$ is the magnetic field gradient along the NV
center axis.

The nanodiamond is trapped inside an optical cavity to pre-cool its
center-of-mass motion to the ground state through the cavity-assisted
cooling as has been demonstrated for other mechanical systems \cite%
{mechanicalground}. The heating of the mechanical mode is negligible
compared with the cavity-induced cooling rate as the $Q$-factor for the
center-of-mass oscillation of an optically levitated particle in vacuum is
very high \cite{Chang10}. Note that the center-mass oscillation is not
subject to the intrinsic dissipation of the material as it is decoupled from
all the other mechanical modes and can be cooled to the ground state even if
the diamond itself (its internal modes) is at room temperature. An
alternative way for the ground state cooling of the center-of-mass mode,
although not demonstrated yet in experiments, is to use a combination of
optical pumping of the NV\ spin state and fast exchange between the spin and
the motional excitations \cite{Rabl09}. In this case, we do not need any
optical cavity, which can further simplify the experimental setup.

\section{Preparation and detection of Fock states}
In order to
prepare the Fock states, we first cool the mechanical mode to the ground
state. The NV spin is initially set to the state $|0\rangle $, which is
decoupled from the mechanical mode during the cooling. Initialization and
single shot detection of the NV spin have been well accomplished
experimentally \cite{Robledo11}. We assume that the NV\ center is at a
position with zero magnetic field and a large field gradient. We apply a
microwave drive with the Hamiltonian $H_{drive}=\hbar (\Omega _{\mathrm{NV}%
,+1}e^{i\omega _{l+}t}|0\rangle \langle +1|+\Omega _{\mathrm{NV}%
,-1}e^{i\omega _{l-}t}|0\rangle \langle -1|+h.c.)/2$ and set the Rabi
frequency $\Omega _{\mathrm{NV},\pm 1}=\Omega _{\mathrm{NV}}$ and the
detuning $\Delta _{\pm }\equiv \omega _{l\pm }-\omega _{\pm 1}=\Delta $.
With $\Delta \gg |\Omega _{\mathrm{NV}}|$, we adiabatically eliminate the
level $|0\rangle $ and get the following effective Hamiltonian
\begin{equation}
\begin{aligned} H_{e} =\hbar \omega_m a_m^\dagger a_m +\hbar\Omega\sigma_z +
\hbar \lambda (\sigma_+ + \sigma_-) (a_m +a_m^\dagger), \end{aligned}
\label{eq:Heff1}
\end{equation}%
where $\Omega =|\Omega _{\mathrm{NV}}|^{2}/4\Delta $, $\sigma _{z}=|+\rangle
\langle +|-|-\rangle \langle -|$, $\sigma _{+}=|+\rangle \langle -|$, $%
\sigma _{-}=|-\rangle \langle +|$, and we have defined the new basis states $%
|+\rangle =(|+1\rangle +|-1\rangle )/\sqrt{2}$, $|-\rangle =(|+1\rangle
-|-1\rangle )/\sqrt{2}$. In the limit $\lambda \ll \omega _{m}$, we set $%
\Omega =\omega _{m}/2$ and use the rotating wave approximation to get an
effective interaction Hamiltonian between the mechanical mode and the NV
center spin, with the form
$$H_{JC}=\hbar \lambda \sigma _{+}a_{m}+h.c..$$
This represents the standard Jaynes-Cummings(J-C) coupling Hamiltonian.
Similarly, if we set $\Omega =-\omega _{m}/2$, the anti J-C Hamiltonian can
be realized with
$$H_{aJC}=\hbar \lambda \sigma _{+}a_{m}^{\dagger }+h.c..$$

\begin{figure}[tbph]
\centering
\includegraphics[width=8.5cm]{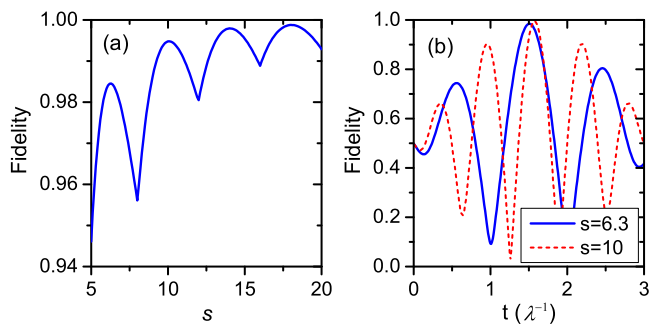}
\caption{(Color online) Fidelity of creating the phonon number superpostion state $%
(|0\rangle _{m}+i|1\rangle _{m})/\protect\sqrt{2}$ by coherent state
transfer between the NV spin and the mechanical mode. Fidelities higher than
99\% can be achieved. (a) The peak fidelity as a function of the parameter $%
s=\protect\omega _{m}/\protect\lambda $. (b) The fidelity as a function of
the interaction time for two different parameters ($s=6.3,10$). }
\end{figure}

Arbitrary Fock states and their superpositions can be prepared with a
combination of J-C and anti J-C coupling Hamiltonians. For example, to
generate the Fock state $|2\rangle _{m}$, we initialize the state to $%
|+\rangle |0\rangle _{m}$, turn on the J-C coupling for a duration $%
t_{1}=\pi /(2\lambda )$ to get $|-\rangle |1\rangle _{m}$, and then turn on
the anti J-C coupling for a duration $t_{2}=t_{1}/\sqrt{2}$ to get $%
|+\rangle |2\rangle _{m}$. The Fock state with arbitrary phonon number $%
n_{m} $ can be generated by repeating the above two basic steps, and the
interaction time is $t_{i}=t_{1}/\sqrt{i}$ for the $i$th step \cite%
{Meekhof96}. Superpositions of different Fock states can also be generated.
For instance, if we initialize the state to $(c_{0}|+\rangle +c_{1}|-\rangle
)\otimes |0\rangle _{m}/\sqrt{2}$ through a microwave with arbitrary
coefficients $c_{0},c_{1}$, and turn on the J-C coupling for a duration $%
t_{1}$, we get the superposition state $|-\rangle \otimes (c_{1}|0\rangle
_{m}+ic_{0}|1\rangle _{m})/\sqrt{2}$. In Fig. 2a, we plot the fidelity of
the mechanical state as a function of the parameter $s=\omega _{m}/\lambda $
using the full Hamiltonian with rotating wave approximation. The fidelity
oscillates with many local maxima and the envelope approaches unity when $%
s\gg 1$. In practice, we have a very high fidelity already by setting $s$ at
the local maxima such as $6.3$ or$10.0$. Using the optical cavity, the Fock
state $|n_{m}\rangle _{m}$ of mechanical mode can also be mapped to the
corresponding Fock state of the output light field \cite{Yin11}.

The effective Hamiltonian for the spin-phonon coupling takes the form
$$H_{QND}=\hbar \chi \sigma _{z}a_{m}^{\dagger }a_{m},$$
with $\chi =4\Omega
\lambda ^{2}/(4\Omega ^{2}-\omega _{m}^{2})$ when the detuning $||\Omega
|-\omega _{m}/2|\gg \lambda $. The Hamiltonian $H_{QND}$ can be used for a
quantum non-demolition measurement(QND) measurement of the phonon number: we
prepare the NV center spin in a superposition state $|+\rangle +e^{i\phi
}|-\rangle )/\sqrt{2}$, and the phase $\phi $ evolves by $\phi (t)=\phi
_{0}+2\chi n_{m}t$, where $n_{m}=a_{m}^{\dagger }a_{m}$ denotes the phonon
number. Through a measurement of the phase change, one can detect the phonon
number.

The preparation and detection of the Fock states can all be done
within the spin coherence time. Let us estimate the typical parameters. A
large magnetic field gradient can be generated by moving the nanodiamond
close to a magnetic tip.
A large field gradient up to $4\times 10^{7}$~T/m
has been reported in 2006 near the write head of a magnetic disk drive \cite%
{Tsang06}.
In magnetic resonance force microscopy systems, the gradient in the
order of $10^6$~T/m has been realized in 2007 \cite{Mamin07}.
Here we take the gradient $G=10^{5}$~T/m and get the coupling $%
\lambda \simeq 2\pi \times 52$~kHz for a nanodiamond with the diameter $d=30$%
~nm in an optical trap with a trapping frequency $\omega _{m}=2\pi \times
0.5 $~MHz. The Fock states and their superpositions can then be generated
with a time scale $1/\lambda $ about a few $\mu $s, and the QND detection
rate $2|\chi |\sim 2\pi \times 25$~kHz with the detuning $||\Omega |-\omega
_{m}/2|\sim 5\lambda $. The NV electron spin dephasing time over $1.8$ ms
has been observed at room temperature \cite{Bala09}, which is long compared
with the Fock state preparation time $1/\lambda $ and the detection time $%
1/\left( 2|\chi |\right) $. The threshold gradient is $2\times 10^3$ T/m for
the trap frequency $0.5$ MHz.

\begin{figure}[t]
\includegraphics[width=8.5cm]{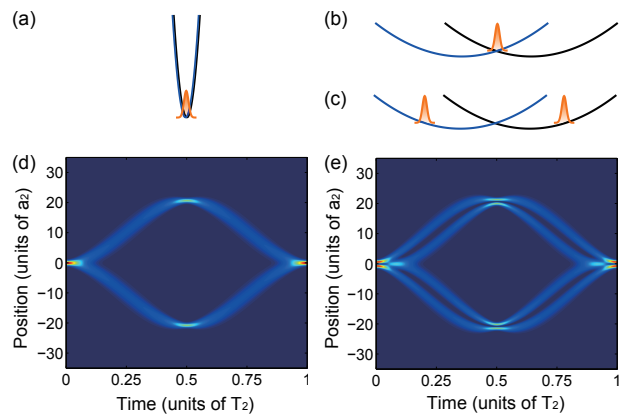}
\caption{ (Color online) (a) A nanodiamond with $d=30$~nm is confined tightly by a $100$kHz
frequency optical tweezer in a magnetic field with a large gradient $%
G=4\times 10^{4}$~T/m. Its NV center is prepared in a state $|0\rangle $.
(b) The power of the optical tweezer is suddenly reduced to $20$ kHz, and NV
center is changed to a superposition state $(|+1\rangle +|-1\rangle )/%
\protect\sqrt{2}$, while the magnetic gradient is the same. As a result, the
trap centers for different electron spins are separated. (c) The nanodiamond
becomes in a spatial superposition state as the state evolves in time.
(d) and (e) show the time evolution of the probability distribution of the nanodiamond
after the trapping frequency and the NV center state are suddenly changed.
The mechanical state is initially  $|0\rangle _{m,100\mathtt{kHz}}$ (d) or
 $|1\rangle_{m,100\mathtt{kHz}}$ (e) in the high frequency trap. $a_{2}=
\protect\sqrt{\hbar /2m\protect\omega _{m2}}=0.092$~nm, and $T_{2}=50\protect
\mu $s.}
\label{fig:distance}\centering
\end{figure}

\section{Generation and detection of large spatial superposition states.}
 To prepare spatial quantum superposition state, we need to generate
quantum superposition of the nanodiamond at distinct locations. Without the
microwave driving, the spin-mechanical coupling Hamiltonian takes the form
\begin{equation}
H=\hbar \omega _{m}a_{m}^{\dagger }a_{m}+\hbar \lambda
S_{z}(a_{m}+a_{m}^{\dagger }).  \label{eq:Hlow}
\end{equation}%
The mechanical mode is initialized to the vacuum state $|0\rangle_m $ (or
a Fock state $|n_{m}\rangle_m $) in a strong trap with the trapping frequency $%
\omega _{m0}$ and the NV center spin is prepared in the state $|0\rangle $.
Although the ground state cooling is most effective in a strong trap, to
generate large spatial separation of the wave packets it is better to first
lower the trap frequency by tuning the laser intensity for the optical trap.
While it is possible to lower the trap frequency through an adiabatic sweep
to keep the phonon state unchanged, a more effective way is to use a
non-adiabatic state-preserving sweep \cite{Chen10}, which allows arbitrarily
short sweeping time.
We denote $|n_{m}\rangle _{m1}$ as the mechanical state in the lower
frequency $\omega _{m1}$. We then apply an impulsive microwave pulse to
suddenly change the NV spin to the state $(|+1\rangle +|-1\rangle )/\sqrt{2}$
and simultaneously decrease the trap frequency to $\omega _{m2}\leq \omega
_{m1}$. The evolution of the system state under the Hamiltonian (2) then
automatically split the wave packet for the center-of-mass motion of the
nanodiamond (see the illustration in Fig. 3). The splitting attains the
maximum at time $T_{2}/2=\pi /\omega _{m2}$, where the maximum distance of
the two wave packets in the superposition state is $D_{m}=8\lambda
a_{2}/\omega _{m2}=4g_{s}\mu _{B}G/(m\omega _{m2}^{2})$, where $a_{2}=\sqrt{%
\hbar /2m\omega _{m2}}$. At this moment, the system state is
\begin{equation}\label{eq:displacement}
 |\Psi_{S}\rangle =(|+1\rangle |D_{m}/2\rangle _{n_{m}}+|-1\rangle
|-D_{m}/2\rangle _{n_{m}})/\sqrt{2},
\end{equation}
where $|\pm D_{m}/2\rangle
_{n_{m}}\equiv (-1)^{a_{m}^{\dagger }a_{m}} e^{ \pm
D_{m}(a_{m}^{\dagger }-a_{m})/4a_{2}} \left\vert n_{m}\right\rangle
_{1}$ is the displaced Fock state (or coherent states when $n_{m}=0$).
This is just the entangled spatial superposition state, as discussed in Appendix \ref{app:1}. In
Fig. 3 (d) and (e), we show the evolution of the splitting of the spatial
wave packets for the nanodiamond under the initial vacuum $|0\rangle_m$
or Fock state $|1\rangle_m$. The maximum distance $D_{m}$ is plotted in
Fig. 4 versus trap frequency, magnetic field gradient, and diameter $d$ of
the nanodiamond, and superposition states with separation $D_{m}$ comparable
to or larger than the diameter $d$ is achievable under realistic
experimental conditions.

To transform the entangled cat state $|\Psi
_{S}\rangle $ to the standard cat state $\left\vert \psi _{\pm
}\right\rangle _{n_{m}}\equiv (|D_{m}/2\rangle _{n_{m}}\pm |-D_{m}/2\rangle
_{n_{m}})/\sqrt{2}$, we need to apply a disentangling operation to
conditionally flip the NV spin using displacement of the diamond as the
control qubit. This can be achieved as different displacements of the
wavepacket induce relative energy shifts of the spin levels due to the
applied magnetic field gradient. As an estimate, for the example we
considered in Fig. 5 (with a $30$nm-diameter diamond in a $20$ kHz trap
under a magnetic gradient of $3\times 10^{4}$ T/m), the spin energy
splitting is about $2.4$ MHz between the $|+1\rangle |D_{m}/2\rangle _{n_{m}}
$ and $|-1\rangle |-D_{m}/2\rangle _{n_{m}}$ components, which is much
larger than the typical transition linewidth of the NV spin (in the order of
kHz). So we can apply first an impulsive microwave pulse to transfer the
component state $|+1\rangle |D_{m}/2\rangle _{n_{m}}$ to $|0\rangle
|D_{m}/2\rangle _{n_{m}}$ without affecting $|-1\rangle |-D_{m}/2\rangle
_{n_{m}}$ and then another pulse to transfer $|-1\rangle |-D_{m}/2\rangle
_{n_{m}}$ to $\pm |0\rangle |-D_{m}/2\rangle _{n_{m}}$. After the two
pulses, the spin state gets disentangled and the position of the diamond is
prepared in the quantum superposition state $\left\vert \psi _{\pm
}\right\rangle _{n_{m}}$.

\begin{figure}[t]
\centering
\includegraphics[width=8.5cm]{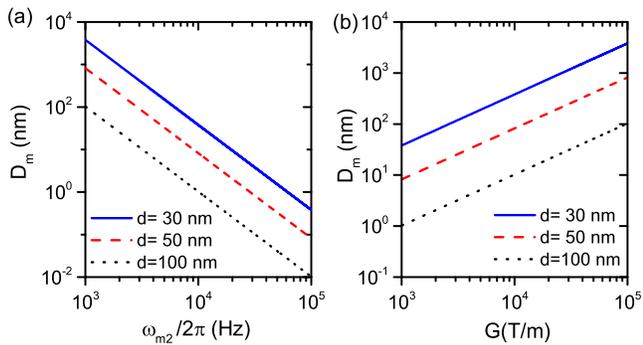}
\caption{(Color online) (a) Maximum spatial separation $D_{m}$ of the superposition state
as a function of trap frequency $\protect\omega _{m2}$ when the magnetic
gradient is $10^{5}$ T/m. (b) Maximum spatial separation $D_{m}$ as a
function of the magnetic gradient $G$ when the trapping frequency is 1 kHz.
Macroscopic superposition states with separation larger than the size of the
particle can be achieved with a moderate magnetic gradient. }
\label{fig:distance1}
\end{figure}

To detect spatial superposition state, we can turn off the optical trap and
let the spatial wave function freely evolve for some time $t$. The split
wave packets will interference just like the Young's double slit experiment.
The period of the interference pattern is $\Delta z=2\pi \hbar t/(mD_{m})$ (Appendix \ref{app:2}).
As an estimation of typical parameters, we take $%
\omega _{m1}=\omega _{m2}=2\pi \times 20$ kHz, $d=30$ nm, and magnetic field
gradient $3\times 10^{4}$ T/m. The spin-phonon coupling rate $\lambda \simeq
2\pi \times 77$ kHz and the maximum distance $D_{m}\simeq 31 a_2$. The
preparing time of sperposition state is about $25$ $\mu $s, which is much
less than the coherence time of the NV spin. For the time of flight
measurement after turn-off of the trap, we see the interference pattern with
a period of $47$ nm after $t=10$~ms, as shown in Fig. 5, which is large
enough to be spatially resolved \cite{li10, li11,Gieseler12}.
In experiments, the initial state is the thermal state with
$\langle n_m \rangle \ll 1$, which can be approximated by the vacuum and the first Fock state.
By combining the interference pattern of vacuum and $|1\rangle_m$ state with its thermal weight,
we can get the interference pattern for $n_m \ll 1$ thermal state.
As show in Fig. \ref{fig:interference1}, because the weight of $|1\rangle_m$ is $n_m \ll 1$,
the final interference patter is very close to the vacuum one. In other words, the
interference pattern is robust for the  thermal phonon number.

\begin{figure}[t]
\includegraphics[width=8.5cm]{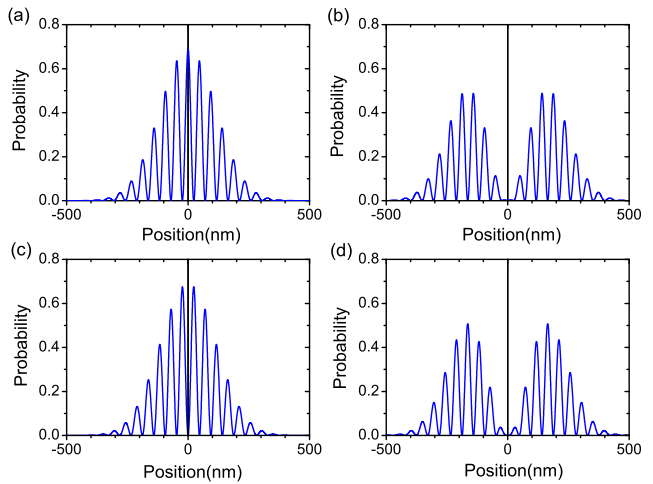}
\caption{(Color online) Spatial interference patterns for a 30~nm nano-diamond after 10~ms
of free expansion. The nano-diamond is initially prepared in the vacuum
state $|0 \rangle_{m}$ or the 1-phonon state $|1 \rangle_{m}$ of a 20 kHz
trap. The magnetic gradient is $3 \times 10^4$T/m. Before the trap is turned
off, the center of mass of the nano-diamond is prepared in (a) $|\protect\psi%
_+\rangle_0$, (b) $|\protect\psi_+\rangle_1$, (c) $|\protect\psi_-\rangle_0$%
, and (d) $|\protect\psi_-\rangle_1$. }
\label{fig:interference}
\centering
\end{figure}

\begin{figure}[t]
\includegraphics[width=8.5cm]{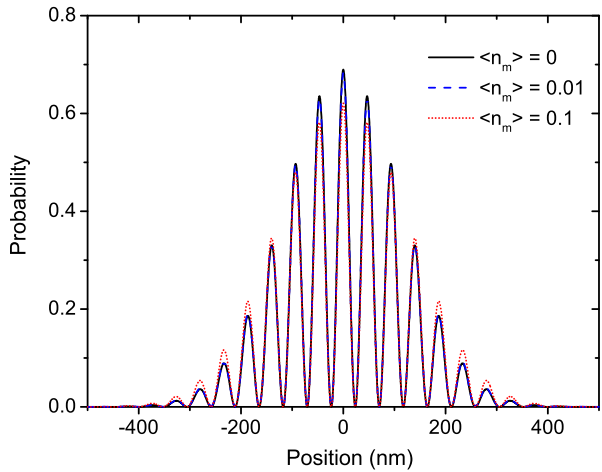}
\caption{(Color online) Spatial interference patterns for a 30~nm nano-diamond after 10~ms
of free expansion. The magnetic gradient is $3 \times 10^4$T/m.
 The nano-diamond is initially prepared in the thermal state of a 20 kHz
trap. Before the trap is turned off, the center of mass of the nano-diamond is
prepared in the spatial quantum superpostion state with initial thermal phonon number $\langle n_m\rangle=0$, $0.01$, $0.1$. }
\label{fig:interference1}
\centering

\end{figure}

\section{Discussion and conclusion}
Finally, we briefly mention the source of decoherence for preparation of the
spatial quantum superposition state. The decoherence of the cat state by
photon scattering is negligible during the time-of-flight measurement as the
laser is turned off. The main source of decoherence includes background gas
collision and black-body radiation. Using the formula
$$\gamma_s = 4\pi\sqrt{2\pi}Pd^2/(\sqrt{3}\bar{v}m_a)$$
in Ref. \cite{Isart11}, we estimate that the decoherence rate due to the background gas collision
is about $8$ Hz under the condition that the vacuum pressure $P\sim 10^{-11}$
Torr, the mass of the molecule $m_a=4.83\times 10^{-26}$ kg, and the gas temperature $T_{b}\sim 4.5$ K. \
Here $\bar{v}=\sqrt{8k_B T/(\pi m_a)}$ is the mean velocity of molcule, $k_B$ is Boltzmann constant. This rate is small compared
with the experimental time scale of $10$~ms. As the internal temperature $%
T_{i}$ of the diamond is typically much higher than the background gas
temperature $T_{b}$, the black-body radiation is dominated by the thermal
photon emission from the diamond, which is $\gamma_{bb,e}= (2\pi^5/189)cd^3(k_B T_i/\hbar c)^6\text{Im}[(\epsilon_{bb}-1)/
(\epsilon_{bb}+2)] z^2$ \cite{Isart11}. Here $c$ is the light speed constant,
$\epsilon_{bb}$ is the permittivity of diamond for the BB radiation, and $z$ is the interference width. The frequency of the thermal photon is in
the order of THz. Compared with the silica sphere considered in \cite%
{Isart11}, the emission rate of THz photons is smaller by about two orders
of magnitude for the diamond. Even with $T_{i}$ at the room temperature, the
decoherence rate $\gamma _{bb}$ due to the thermal photon emission is
estimated to be only $3$ Hz for our proposed scheme. In addition, with the
built-in electronic states of the NV\ center, the internal temperature of
the diamond can be reduced through laser cooling \cite{cooling}. As $\gamma
_{bb}\propto T_{i}^{6}$, this allows further significant suppression of the
decoherence rate. For the current discussion of $30$ nm particle in the $0.5$MHz trap,
the gravity induced decoherence is estimated to be in the order
of $10^{-62}$ Hz for the superposition of Fock states \cite{Blencowe12}, and $10^{-7}$ Hz for spatial
superposition states with displacement $30 $ nm \cite{Penrose1996}, and can be neglected.
Our proposed scheme is more relevant for testing objective collapse models \cite{Isart11} .

We note that the same scheme could also apply to other optically levitated
nano-particles with build-in electron spins, such as ${}^{28}$Si
nano-particles with donor spins \cite{Steger10} or nano-crystals doped with
rare-earth ions \cite{Titte10}. This scheme thus opens up the possibility to
observe large quantum superpositions for mesoscopic objects under realistic
experimental conditions and to test predictions of quantum mechanics in
unexplored regions.

We thank J.-N. Zhang, K. C. Fong, and S. Kheifets for discussions. This work
was funded by the NBRPC (973 Program) 2011CBA00300 (2011CBA00302), NNSFC
11105136, 61033001. LMD acknowledge support from the IARPA MUSIQC program,
the ARO and the AFOSR MURI program. TL and XZ were supported by NSF
Nanoscale Science and Engineering Center (CMMI-0751621).

\appendix

\section{Derivation of entangled spatial superposition state}
\label{app:1}

The Hamiltonian (2) in rotating frame with $H_0=\hbar \omega_m a_m^\dagger
a_m$ reads $H_I= \hbar S_z \lambda (ae^{-i\omega_m t} + h.c.)$. The unitary
operator in the rotating frame is $U_I = \mathcal{T}\exp [ -\frac{i}{\hbar}
\int H_I(t) \mathrm{d} t]$, which can be solved with Magnus expansion $U_I =
\exp [\sum_k \Omega_k (t)]$. Here
\begin{equation}
\begin{aligned}
  \Omega_1= &-\frac{i}{\hbar} \int_0^t H_I
(t_1) \mathrm{d}t_1\\
=&\frac{S_z\lambda}{\omega_m}[(e^{-i\omega_m t} -1)a -
(e^{i\omega_m t}-1) a^\dagger],
\end{aligned}
\end{equation}
and
\begin{equation}
 \begin{aligned}
  \Omega_2 =& -\frac{1}{2} \int_0^t
\int_0^{t_1} [H_I (t_1), H_I (t_2)] \mathrm{d}t_1 \mathrm{d} t_2 \\
 =& i\lambda^2
(\frac{t}{\omega_m} - \frac{\sin (\omega_m t)}{\omega_m^2}).
\end{aligned}
\end{equation}
For $k\geq 3$, $%
\Omega_k$ are always equal to zero. Neglecting the global phase $\Omega_2$,
we get $U_I = \exp [ \alpha(t) a - \alpha(t)^* a^\dagger ]$, where $\alpha
(t) =S_z \lambda( e^{-i\omega_m t}-1)  /\omega_m$. Back to original frame,
we get $U_I^{\prime}=e^{ -i\omega_m a^\dagger_m a_m t} U_I$. Specifically, at the
time of half period with trap frequency $\omega_m=\omega_{m_2}$, we find
that $U_I^{\prime}=(-1)^{ a_m^\dagger a_m} \exp [S_zD_m (a_m^\dagger -a_m)/4a_2]$,
where $D_m =4 g_s \mu_B G /(m\omega_{m_2}^2 )$, and $a_2= \sqrt{%
\hbar/(2m\omega_{m_2})}$. The entangled spatial state can be calculated as $%
|\Psi_S\rangle = U_I^{\prime }|n_m\rangle_{m1}$.

\section{Period of interference pattern}
\label{app:2}
We suppose that the superpostion of displacement state is $|\psi_+\rangle_0=
[\varphi_{0L}(z)+ \varphi_{0R} (z)]/\sqrt{2}$, where $\varphi_{0R} (z) =
\frac{\sqrt{\beta}}{\pi^{1/4}}e^{-\beta^2(z-b)^2/2}$, $\varphi_{0L} (z) =
\frac{\sqrt{\beta}}{\pi^{1/4}}e^{-\beta^2(z+b)^2/2}$, and $\beta = \sqrt{%
m\omega_2/\hbar}=1/(\sqrt{2}a_2)$. $b$ is the displacement of the original
wavefunction. After the Fourier transfer , the right branch wave function
becomes
$$\varphi_{0L}(k) = \frac{1}{\pi^{1/4}\sqrt{\beta}}
e^{-ibk-(k^2/2\beta^2)}.$$
 The Fourier transfer of the left branch of the
wavefunction can be calculated with the similar method. Time evolution of
the wavefunction is
$$\varphi_0(z,t)= \frac{1}{\sqrt{2\pi}}
\int_{-\infty}^{+\infty} d k \varphi(k) e^{i(kz-\omega t)},$$
where $\omega=\hbar k^2/2m$. We suppose that $z$ in the unit of $1/\beta$, $t$ is in the unit of
$2m/\hbar \beta^2$, and $k$ is in the unit of $\beta$. The probability
distribution of the wave function is
\begin{eqnarray*}
\varphi_0^* (z,t) \varphi_0 (z,t)&=& \frac{1}{2\sqrt{\pi(1+4t^2)}}\big( e^{-%
\frac{(z-b)^2}{1+4t^2}} + e^{-\frac{(z+b)^2}{1+4t^2}}  \notag \\
&+& 2 e^{-\frac{z^2+b^2}{1+4t^2}} \cos (\frac{4bzt}{1+4t^2})\big).
\end{eqnarray*}
The interference period is $\Delta z = 2\pi (1+4t^2)/(4bt)$. For long time
limit $bt \gg 1$, we have $\Delta z = 2\pi t/b$. In our paper, we have $%
D_m=2b$. Therefore, in the S.I. unit the interference period is $\Delta z =
2\pi \hbar t/(mD_m)$.

\end{document}